\documentstyle[aps,preprint,epsf]{revtex}
\begin{document}

\title{Diffusion and multifractality at the metal-insulator
  transition}

\author{Bodo Huckestein and Rochus Klesse}

\address{Institut f\"ur Theoretische Physik, Universit\"at zu
  K\"oln, D-50937 K\"oln, Germany}  

\date{\today}

\maketitle

\begin{abstract}
  
  We review the time evolution of wavepackets at the metal-insulator
  transition in two- and three-dimensional disordered systems. The
  importance of scale invariance and multifractal eigenfunction
  fluctuations is stressed. The implications of the frequency- and
  wavevector-dependence of the diffusion coefficient are compared with
  the results of numerical simulations. We argue that network models
  are particularly suited for the investigation of the dynamics of
  disordered systems.

\end{abstract}


\section{Introduction}
\label{sec:intro}

The time evolution of wavepackets reflects clearly the localization
properties of the eigenstates of a system. While in infinite metallic
systems wavepackets spread indefinitely and take on a Gaussian shape
for long times, diffusion is absent in localized systems and
wavepackets stay in a finite region around their starting point. In
fact, the consideration of the diffusion of wavepackets has been the
starting point of the theory of localization in random media
\cite{And58}. At the mobility edge between metallic and insulating
behavior the time evolution of wavepackets will be intermediate
between the two extreme cases discussed above. The dynamics of
wavepackets can be described by a frequency- and wavevector-dependent
diffusion coefficient $D(q,\omega)$. Its behavior has previously been
studied in detail \cite{Cha90,HS94,BHS96}. The diffusion coefficient
exhibits two characteristic features. First, from the scale invariance
of the conductance \cite{AALR79} follows that the conductivity and via
the Einstein relation the diffusion coefficient are scale dependent in
higher than two dimensions. Thus only in two dimensions there exists a
finite limit of $D(q,\omega)$ as frequency and wavevector tend to
zero. Secondly, on short distances and frequencies the diffusion
coefficient reflects the multifractal correlations within single
eigenfunctions of the system \cite{Weg76}.

In this paper we focus on the time evolution of wavepackets. While the
same information is encoded in the frequency- and
wavepacket-dependence of the diffusion coefficient, we concentrate on
the time dependence of characteristics of wavepackets like their
return probability, their moments, and their shape. Our purpose in
doing so is twofold. On the one hand elucidate these discussions some
of the crucial differences between time evolution in the diffusive
regime and at the mobility edge. On the other hand can the time
evolution of wavepackets directly been studied in numerical
simulations of network models, allowing for a check of predictions of
scaling theory.

The paper is organized as follows: we first summarize the scaling form
of the diffusion coefficient $D(q,\omega)$. Next we present the time
dependence of the moments, return probability, and shape of
wavepackets that results from this scaling form. And finally, we
compare these predictions with numerical results obtain for network
models.

\section{Scaling form of the diffusion coefficient}
\label{sec:scale}

In this section we will briefly review known results about the scaling
form of the diffusion coefficient. For a more complete discussion see,
e.g., refs. \cite{Cha90,BHS96}. 

Consider the time evolution of a wavepacket that at time $t=0$ is
localized as a $\delta$-function at the origin. Due to conservation of
the probability the Fourier (Laplace) transform of its probability
density with respect to space and time has of the form
\begin{equation}
  \label{pqw}
  P(q,\omega) = \frac{1}{-i\omega+D(q,\omega)q^2}.
\end{equation}
A scaling form for $D(q,\omega)$ that is consistent the conductance
being scale invariant at the metal-insulator transition is
\begin{equation} 
D(q,\omega )=\frac{\xi^{2-d}}{\hbar\rho}F\left(\xi/L_\omega,\xi q\right),
\end{equation} 
where $\xi$ is the localization length, $\rho=\rho (E)$ the density of
states near the critical energy $E=E_{c}$, $F$ is a scaling
function, and
\begin{equation}
\label{lw}
L_{\omega } := (\rho \hbar \omega )^{-1/d}=L(\hbar\omega /\Delta )^{-1/d},
\end{equation}
is a third length scale, besides $q^{-1}$ and $\xi$, relevant at the
critical point ($\Delta =(L^{d}\rho )^{-1}$ is the mean level
spacing). At the critical point, where the localization length $\xi$
diverges, the diffusion coefficient can be written as
\begin{equation}
D(q,\omega )=\frac{q^{d-2}}{\hbar\rho }f(qL_{\omega }).
\end{equation}
From various limiting cases the following form of $f(x)$ can be deduced:
\begin{equation} 
\label{fx} 
f(x)= \left\{\begin{array}{r@{\quad:\quad}l} 
c_{\alpha }\cdot x^{(2-d)/d} & x \to 0\\ 
c_{\beta} & \mbox{for intermediate values of }x \\
c_{\gamma  }\cdot x^{-\eta/d} &  x\to \infty
\end{array} \right.
\end{equation} 
The behavior for large $x$ is governed by multifractal density
fluctuations in single eigenfunctions and the correlation exponent
$\eta$ is related to the generalized dimension $D_2$ of the inverse
participation ratio $\eta=d-D_2$ \cite{PJ91}. Numerical support for
this scaling form has previously been obtained from calculations of
the two-particle spectral function and the wavepacket return
probability of real-space and tight-binding models of the quantum Hall
effect \cite{CD88,HS94} as well as three-dimensional tight-binding
models \cite{BHS96}.

\section{Shape of wavepackets}
\label{sec:shape}

The shape of wavepackets $P(r,t)$ can be obtained by transforming
eq.~(\ref{pqw}) back to the space and time domain using the scaling
form~(\ref{fx}). The long time short distance behavior of $P(r,t)$ is
thus dominated by multifractal eigenfunction fluctuations
characterized by the exponent $\eta$. The inverse Laplace transform with
respect to $\omega$ in the limit of $q^d t/\hbar \omega \gg 1$ is
\cite{BHS96} 
\begin{equation}
\label{Pqtasympt}
P(q,t)\sim (q^{d}t)^{-D_{2}/d}.
\end{equation}
The shape of the wavepackets is given by the Fourier transform of this
expression. The short distance behavior, $r^d \ll t/\hbar\rho$, is
thus a power law \cite{KKKG97}
\begin{equation}
  \label{prtasympt}
  P(r,t)\sim t^{-D_2/d}r^{D_2-d}.
\end{equation}
It should be noted that while this is the short distance behavior it
can cover many decades for long times and hence describes completely
the bulk of the wavepacket for long times.

It also follows from eq.~(\ref{Pqtasympt}) that the return probability
$P(r=0,t)$ decreases as a power of time \cite{HS94}
\begin{equation}
  \label{pt}
  P(r=0,t)\sim t^{-D_2/d}.
\end{equation}
We notice that the shape of the wavepacket and hence the return
probability depends of the generalized dimension $D_2$ reflecting the
influence of the multifractal eigenfunction fluctuations. In contrast
to that, the moments of the probability distribution of a wavepacket
$m_{k}(t) = \int d^{d} r\,r^k P(r,t)$ do not depend on $D_2$. From the
scaling form~(\ref{fx}) it follows that $P(q,t)=p(q^dt/\hbar\rho)$
and
\begin{eqnarray}
  m_{k}(t) &=& \int d^d r\,r^k P(r,t) = \Omega_{d}\int_0^\infty dr\,
  r^{d-1+k} \int \frac{d^d q}{(2\pi)^{d}}
  e^{-i\mathbf{q}\cdot\mathbf{r}} 
  P(q,t),\\
  &=& t^{k/d} \Omega_d\int_0^\infty dy\, y^{d-1+k} \int \frac{d^d
  x}{(2\pi)^{d}} e^{-i\mathbf{x}\cdot\mathbf{y}} p(x),
\end{eqnarray}
where $x=q^dt/\hbar \omega$ and $\Omega_d$ is the surface area of the
$d$-dimensional unit sphere. The moments of the wavepacket are
integral aspects of the shape of the wavepacket that are not dominated
by its short distance behavior. Instead due to scale invariance they
scale like $t^{k/d}$.

The behavior of wavepackets at the mobility edge should be contrasted
to the well-known metallic behavior of asymptotically Gaussian
wavepackets
\begin{equation}
  \label{gauss}
  P(r,t)=\frac{e^{-r^{2}/(4Dt)}}{(4\pi Dt)^{d/2}},
\end{equation}
with constant diffusion coefficient $D$ and moments $m_k(t)\sim
t^{k/2}$. The exponents at the critical point differ from these
results in that a) the space dimension $d$ is replaced by the
generalized dimension $D_2$ due to the multifractal fluctuations and
b) the factor $2$ is replaced by $d$ reflecting the scale invariance
of the conductance that gives rise to factors with exponents $d-2$ in
the critical case.

It is further instructive to compare our results for the wavepacket
dynamics at the mobility edge of a disordered system to those obtained
for general quantum-mechanical systems. Under quite general
assumptions it was shown that the return probability decays like
$t^{-\tilde{D}_2}$, where $\tilde{D}_2$ is the generalized dimension
of the spectral measure or the energy dependent local density of
states \cite{KPG92}. In the other hand, the exponent of the spatial
decay of a wavepacket is given by $D_2-d$, where $D_2$ is the
generalized dimension of the spatial dependence of the local density
of states discussed above \cite{KKKG97}. Finally, in the absence of
multiscaling the moments of a wavepacket grow like $t^{k\beta}$ with
$\beta=\tilde{D}_2/D_2$ \cite{KKKG97}. Our results obtained above are
a special case of these more general results. They depend only on a
single exponent $D_2$ since at the mobility edge the generalized
dimensions of the spatial and spectral dependence of the local density
of states are related to each other, $D_q=d\tilde{D}_q$
\cite{HS94,HK96}. The origin of this simplification is the occurrence
of a single relevant length scale $L_\omega$ (eq.~(\ref{lw})),
connecting the energy and length scales.

\section{Numerical results}
\label{sec:numerics}

In order to numerically study the time evolution of wavepackets for a
system defined by a Hamiltonian it is necessary to diagonalize at
least the part of the spectrum containing the support of the
wavepacket and to calculate the corresponding eigenstates.
Alternatively, one can apply the time evolution operator
$\exp(-iHt/\hbar)$ to a wavepacket. Here the problem is that the
kinetic and potential energy terms in the Hamiltonian do not commute
so that a discretization in small timesteps and the use of an
exponential decomposition is necessary \cite{KO95,KO96}. In any case,
the spectrum contains in general extended, critical, and localized
states at the same time. It is therefore difficult to extract the
dynamics at the critical point from any superposition of eigenstates
of a finite system. Both of these problems can be avoided by
investigating network models \cite{Sha82,CC88,KM95}. These models are
defined by a unitary network operator $U$ that can be interpreted as a
time evolution operator \cite{HK96}. The disorder as well as the
energy of the system enters the network operator as parameters. By
tuning the energy and disorder to their critical values the network
operator describes a system at criticality with no contributions from
localized or extended states. A discrete time evolution in such a
system is obtained by repeatedly acting with the operator $U$ on a
network state $\psi$:
\begin{equation}
  \label{time}
  \psi(t) = U^t \psi(0),
\end{equation}
with integer $t$. Since all eigenstates of $U$ are critical it is
possible to start the time evolution with a wavepacket maximally
peaked at a single site. Again, for a Hamiltonian system this is
impossible due to the presence of eigenstates with different
localization properties in the spectrum, limiting the minimum size of
a wavepacket.

In order to investigate the dependence on the dimensionality we study
both two- and three-dimensional systems. The two-dimensional system is
a network model proposed by Chalker and Coddington as a model system
exhibiting the integer quantum Hall effect \cite{CC88}. It has only
localized eigenstates except for states at the single energy in the
center of a Landau level that are critical. The three-dimensional
model consist of a stack of coupled Chalker-Coddington networks
corresponding to a layered system in a strong magnetic field
\cite{CD95}. This system exhibits a true metal-insulator transition
with extended and localized phases separated by a mobility edge with
critical states.

\begin{figure}[htbp]
  \begin{center}
    \leavevmode
    \epsfysize=9cm
    \epsffile{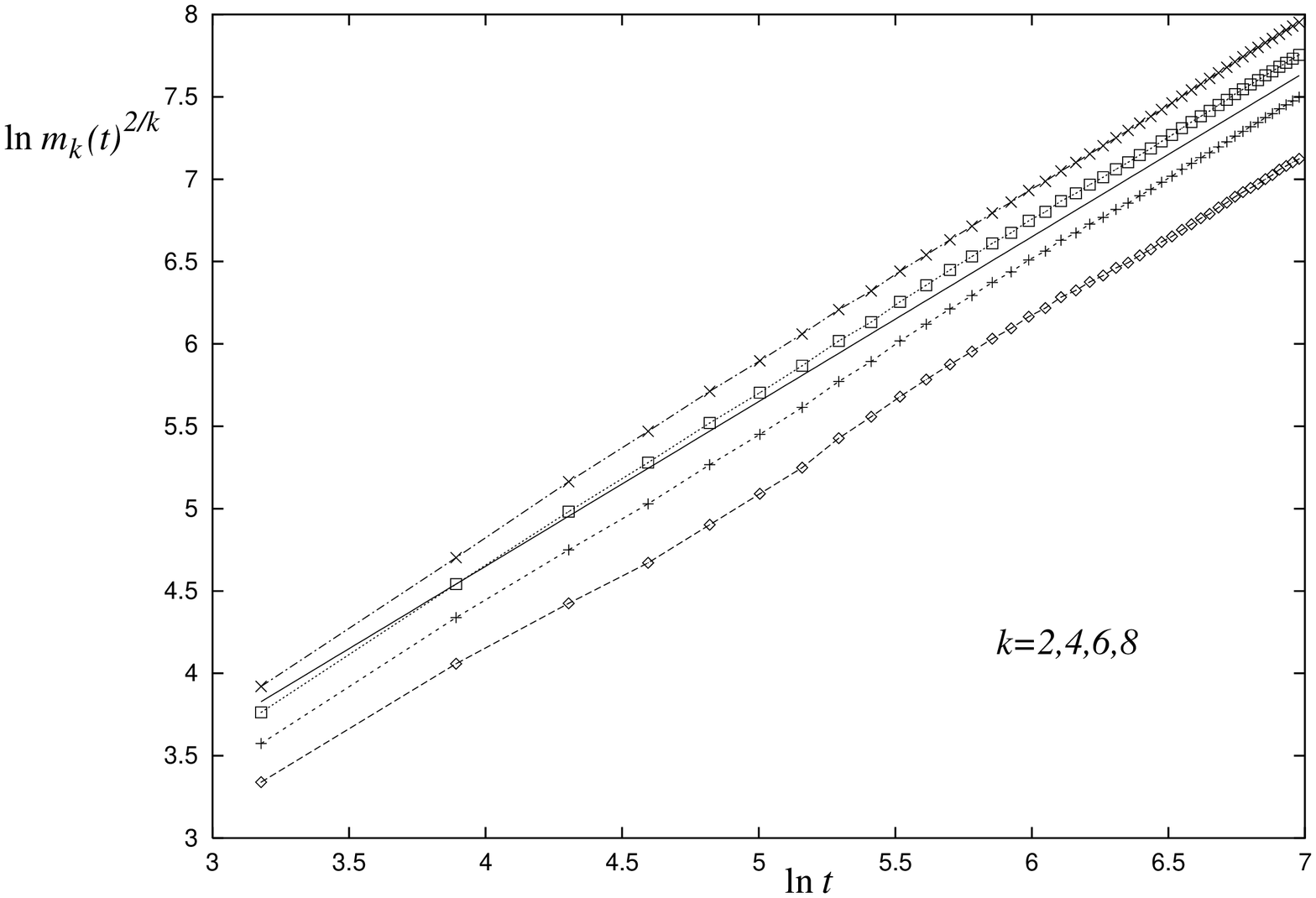}
    \caption{Moments of a wavepacket $m_k(t)^{2/k}$ versus time for a
      quantum Hall system and $k=2,4,6,8$.}
    \label{moment2d}
  \end{center}
\end{figure}

\begin{figure}[htbp]
  \begin{center}
    \leavevmode
    \epsfysize=9cm
    \epsffile{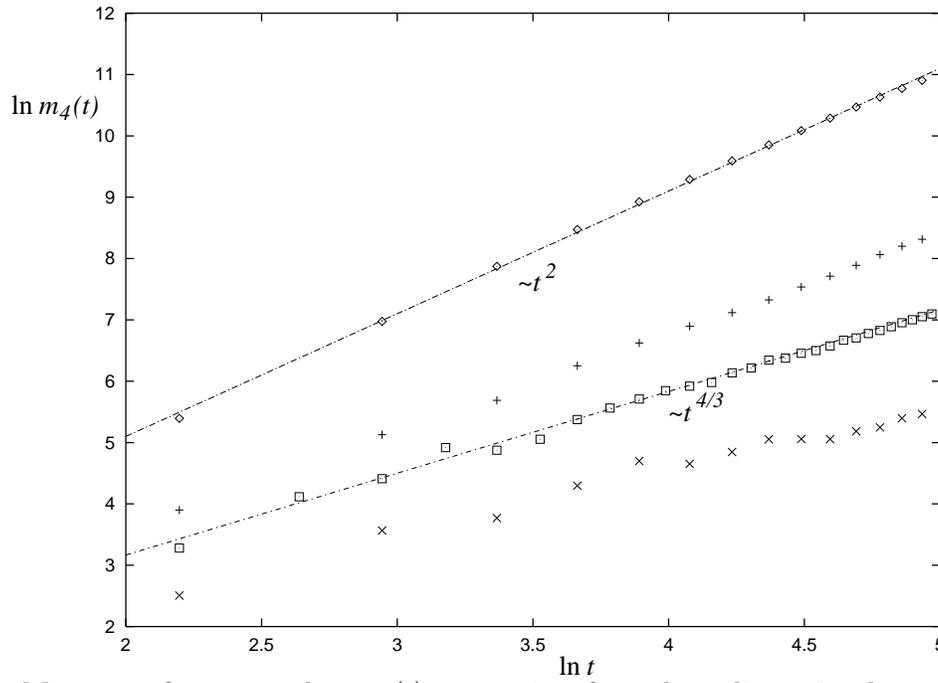}
    \caption{Moments of a wavepacket $m_k(t)$ versus time for a
      three-dimensional system and different energies showing
      diffusive (top), critical (center), and localized behavior
      (bottom).} 
    \label{moment3d}
  \end{center}
\end{figure}

The behavior of the moments of wavepackets is shown in
figs.~(\ref{moment2d}) and (\ref{moment3d}). For the quantum Hall
system in fig.~(\ref{moment2d}) the logarithm of $m_k(t)^2/k$ is
plotted versus $\ln t$ for $k=2,4,6,8$. Note that in $d=2$ the
behavior of the moments is the same in the diffusive as well as the
critical regime. In contrast for the three-dimensional system the
exponent is different in the critical regime, $k/d$, and in the
diffusive regime, $k/2$. This is illustrated in fig.~(\ref{moment3d}),
where $m_4(t)$ is plotted for various energies from the metallic
regime (topmost curve) to the localized regime (bottom curve) with the
critical regime with an exponent of $4/3$ in between.

\begin{figure}[htbp]
  \begin{center}
    \leavevmode
    \epsfysize=9cm
    \epsffile{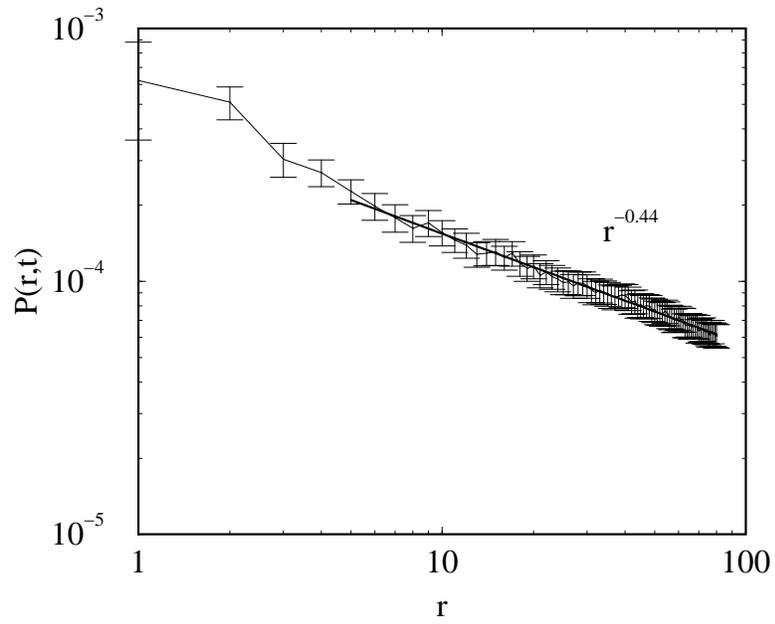}
    \caption{Probability distribution $P(r,t)$ of wavepackets for a
    quantum Hall system. The average over 5 different realizations of
    the disorder is shown. The size of the system is $160\times160$
    and double periodic boundary conditions are imposed.} 
    \label{shape_160}
  \end{center}
\end{figure}

The decay of the return probability has already been studied in
refs.~\cite{HS94,BHS96} finding agreement with the expected exponent
of $D_2/d$. Here we present results for the shape of wave packets in a
two-dimensional network in fig.~(\ref{shape_160}). The topology of the
sample is a torus and the time is chosen such that the power law decay
is observable up to half of the diameter of the system. The observed
exponent of $2-D_2\approx0.44$ agrees well with previous calculations
of $D_2$ from other quantities.

\section{Conclusions}
\label{sec:conc}

The time evolution of wave packets at the mobility edge of disordered
systems differs from the diffusive evolution of wavepackets in good
metals. The wavepackets spread for all times but their shape never
becomes Gaussian. The origin of this behavior is the scale invariance
of the conductance and the multifractal fluctuations of the local
density of states. The latter dominate the short distance, long time
behavior of wavepackets. The return probability decays like
$t^{-D_2/d}$ and the bulk of wavepacket has a power law shape $\sim
r^{D_2-d}$. The moments of a wavepacket depend only weakly on the
multifractal fluctuations in that their growth exponent in time,
$k/d$, does not depend on $D_2$ while the prefactor does. That only a
single multifractal exponent $D_2$ occurs and not one for the spatial
and one for the spectral structure of the local density of states, as
is generally the case, can be interpreted as consequence of the
existence of a single length scale $L_\omega$ connecting energy and
length scales.

The numerical study of the time evolution of wavepackets is most
economically performed for network models compared to models defined
by a Hamiltonian. The network models are defined by a unitary network
operator that provides a natural discrete time evolution. The problem
then reduces to repeated applications of the network operator. The
results of the numerical calculations are in agreement with the
scaling arguments and with numerical calculations based on different
methods.

\section{Acknowledgments}
\label{sec:ack}

One of us (B.H.) gratefully acknowledges previous collaborations with
Ludwig Schweitzer and Tobias Brandes. This work was performed within
the Sonderforschungsbereich 341 of the Deutsche Forschungsgemeinschaft.


\end{document}